\documentclass[cits]{PoS}

\usepackage[intlimits]{amsmath}
\usepackage{amssymb}
\usepackage{bbm}

\title{Baryon axial charges from Chirally Improved fermions - first results}

\ShortTitle{Baryon axial charges from CI fermions}

\author{Georg Engel$^a$, Christof Gattringer$^a$, Leonid Ya. Glozman$^a$, C. B. Lang$^a$, Markus Limmer$^a$,
        \speaker{Daniel Mohler}\hspace{1mm}$^a$ and Andreas Sch\"afer$^b$\\
        \llap{$^a$} Institut f\"ur Physik, FB Theoretische Physik,
        Universit\"at Graz, A-8010 Graz, Austria\\
	\llap{$^b$} Fakult\"at f\"ur Physik, Universit\"at Regensburg, D-93040 Regensburg, Germany\\ 
        E-mail: \email{georg.engel@uni-graz.at},
        \email{christof.gattringer@uni-graz.at},
	\email{leonid.glozman@uni-graz.at},
        \email{christian.lang@uni-graz.at},
        \email{markus.limmer@uni-graz.at},
        \email{daniel.mohler@uni-graz.at},
        \email{andreas.schaefer@physik.uni-regensburg.de}}

\abstract{We present first results from dynamical Chirally Improved (CI) fermion
simulations for the axial charge $G_A$ of various hadrons. We work with
$16^3\times 32$ lattices of spatial extent 2.4 fm and use the variational method
with a suitable basis of Jacobi-smeared interpolators to suppress
contaminations from excited states.}

\FullConference{The XXVII International Symposium on Lattice Field Theory\\
		 July 26-31, 2009\\
		 Peking University, Beijing, China}

\begin{document}

\section{Introduction}

The axial charge of the nucleon, or more precisely the ratio
$G_A(q^2=0)/G_v(q^2=0)$ has been determined to a high
precision from neutron $\beta$ decay, with $G_A(0)/G_v(0)=1.2695(29)$.  In
general, the axial form factor $G_{A,BB^\prime}$ for an octet baryon is given by
\begin{align}
\left<B^\prime|A_\mu(q)|B\right>&=\bar{u}_{B^\prime}(p^\prime)\left(\gamma_\mu\gamma_5G_{A,BB^\prime}(q^2)+\gamma_5q_\mu\frac{G_P(q^2)}{2M_B}\right)u_B(p)\mathrm{e}^{-iq\cdot
  x}\label{ga_def}\ ,
\end{align}
where $G_P$ is the induced pseudoscalar form factor. The axial charge is defined as the value of the axial form factor at zero momentum
transfer $G_{A,BB^\prime}(q^2=0)$. In the following,  we will omit the indices
$B$ and $B^\prime$ when referring to the nucleon. For the nucleon in the
chiral limit, the Goldberger-Treiman relation $G_A=F_\pi g_{\pi NN}/M_N$ connects the axial charge to the pion
decay constant $F_\pi$, the pion-nucleon coupling constant $g_{\pi NN}$ and
the nucleon mass $M_N$. Away from the chiral limit, this relation is still approximately fulfilled. Assuming the conservation of the vector current (which is the case for mass-degenerate light quarks
$m_u=m_d$), the nucleon axial charge is also related
to the polarized quark distributions in the proton: $G_A=\Delta u-\Delta d$
\cite{Sasaki:2003jh}. In an isovector combination, disconnected contributions
cancel, making high-precision lattice computations feasible.

The Chiral Perturbation Theory ($\chi PT$) expressions relevant to the nucleon axial charge have been calculated in
\cite{Beane:Detmold}, where finite volume effects are
taken into consideration. While a recent simulation with domain wall fermions
\cite{Yamazaki:2008py} finds considerable finite volume effects and scaling in $M_\pi L$,
volume effects calculated in $\chi PT$ lead to differing conclusions. Trying
to attribute this difference to excited state contaminations arising from finite separation in
Euclidean time, Tiburzi \cite{Tiburzi:2009zp} estimates the effects of such contaminations
and obtains that they would lead to an over-estimation of $G_A$ rather than
an under-estimation. He also suggests to study $G_A$ using the variational
method. Lattice results for the nucleon axial charge have furthermore been
presented in \cite{Syritsyn:Edwards:Takahashi}. For a recent
review, please refer to the review by Renner \cite{renner_lat09}.

So far, only one group has reported results for the axial couplings of sigma and cascade
hyperons \cite{Lin:2007ap}. The corresponding $\chi$PT calculations can be found in
\cite{Jiang:Tiburzi}. In \cite{Jiang:2009sf} input from
experiment and lattice QCD is used to determine the unknown parameters in the
$\chi PT$ expansion and predict the mass dependence and values
of the axial charges in the chiral limit.

In the next section, we explain the setup for calculations of baryon axial
charges using Chirally Improved (CI) lattice fermions and the variational method. We will then move on and
present results from our calculations of the axial charges of the nucleon and of $\Sigma$ and $\Xi$
hyperons.

\section{Details of our calculational setup}

For our simulations we use CI fermions \cite{Gattringer:CI}, which are approximate
Ginsparg-Wilson fermions based on an expansion of Dirac operator terms on a
hypercube. CI fermions have been tested extensively in quenched calculations \cite{Gattringer:2003qx}
and results for the ground state spectrum of mesons and baryons from dynamical
CI simulations have been presented recently \cite{Gattringer:2008vj}.

Assuming mass-degenerate up- and down quarks it is sufficient to consider
the following current insertion to extract the axial charge
\begin{align}
\label{axial_current}
2A_\mu^3&=A_\mu^u-A_\mu^d\ ,\qquad A_\mu^u=\bar{u}\gamma_\mu\gamma_5 u\ ,\qquad
A_\mu^d=\bar{d}\gamma_\mu\gamma_5 d \ ,
\end{align}
  in which $u$ denotes an up quark and $d$ denotes a down quark. In the
  following, we will show how three-point functions with insertions like those
  in Equation \ref{axial_current} can be evaluated on the lattice.

\begin{figure}[bt]
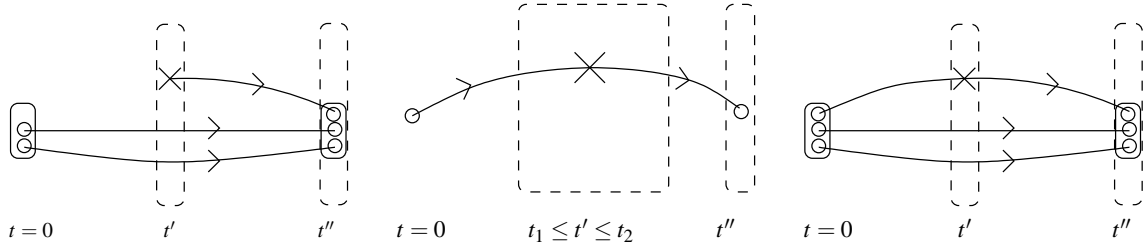

\begin{center}
\input{sequential_a.tex}\hfill
\input{sequential_b.tex}\hfill
\input{sequential_c.tex}  
\caption[Different possibilities for the calculation of sequential quark
  propagators]{Different possibilities for the calculation of sequential quark
  propagators (left and middle). The right-hand side shows an illustration of the full baryon three-point function.
}
\label{sequential_possibilities}
\end{center}
\end{figure}

For our calculations we use
so-called \emph{sequential propagators}. In \cite{Sasaki:2003jh}, two methods for the calculation of sequential quark
propagators are presented. Figure \ref{sequential_possibilities} illustrates
these two approaches. On the left-hand side, sequential sources are built from specific diquark
  propagators, while in the middle, the propagators are calculated for each
  possible insertion separately. An illustration of the
full baryon three-point function is provided on the right-hand side. Which approach is computationally cheaper depends on the physics objective. In our
case, we want to use two different insertions and two widths of smearing for
three different interpolator types. As we are using a rather coarse lattice, a
small number of insertion timeslices should be enough. Therefore, even when just considering the
nucleon case, the second approach using sequential sources from single quark propagators is slightly cheaper. Moreover, these propagators
can subsequently be used for other hadrons and for the calculation of
transition form factors.

To relate lattice operators, which receive a finite renormalization, to their
continuum counterparts, we need to estimate the renormalization factors
$Z_\Gamma$ of the bilinear currents in question. In general, we have to
multiply the lattice result $G_\Gamma^{lat}$ by the appropriate
renormalization factor to obtain values that can be compared with results
extracted from experiments
\begin{align}
G_\Gamma^{phys}=Z_\Gamma G_\Gamma^{lat},
\end{align}
which are typically given in the \emph{modified minimal subtraction} ($\overline{MS}$) renormalization
scheme.

For dynamical CI fermions, these renormalization constants have been estimated
using local bilinear quark field operators in \cite{philipp_renorm}. It would however be useful to have an independent estimation
of these constants from a different method. In the case of the vector
current, one can estimate the constant $Z_V$ by calculating the vector charge
$G_V$ defined in analogy with \eqref{ga_def} via
\begin{align}
\left<B^\prime|V_\mu(q)|B\right>&=\bar{u}_{B^\prime}(p^\prime)\left(\gamma_\mu G_V(q^2)+q_\nu\sigma_{\nu\mu}\frac{G_T(q^2)}{2M_B}\right)u_B(p)\mathrm{e}^{-iq\cdot
  x}\ ,
\end{align}
as $G_V(q^2=0)$. This quantity has to be 1 in the continuum, as it is related to the
electric charge of the proton in the limit of equal quark masses
\cite{Sasaki:2003jh}.

For lattice fermions with exact chiral symmetry, the axial vector renormalization constant $Z_A$ and the vector
renormalization constant $Z_V$ have to be equal. For lattice fermions which
only fulfill the Ginsparg-Wilson relation approximately, there should be small
deviations from this. To obtain an independent estimate of $Z_V$, we use a
ratio of two-point over three-point functions \cite{Burch:2008qx}
\begin{align}
R^{(k)}&=\frac{\sum_l\sum_m\psi_l^{(k)}C(t)_{lm}\psi_m^{(k)}}{\sum_i\sum_j\psi_i^{(k)}T_V(t,t^\prime)_{ij}\psi_j^{(k)}}=Z_V
\ ,
\end{align}
where $C(t)$ is the matrix of two-point correlation functions and
$T_V(t,t^\prime)$ is the matrix of three-point correlators with a vector
insertion. The eigenvectors $\psi$ are the ones obtained from a variational
analysis of $C(t)$. We then compare with the preliminary estimates from \cite{philipp_renorm}. The ensemble names are according to
\cite{Gattringer:2008vj}, where details of the run parameters are
provided. For runs A and B the two methods agree within $2-3\%$. While a
determination using local quark bilinears yields 0.818(2) for run A and
0.826(1) for run B, we find values for $Z_V$ of 0.803(2) and 0.792(2)
respectively. For run C there is a rather large discrepancy and the method of
\cite{philipp_renorm} leads to a value of 0.829(1) while we obtain 0.77(1). Notice also that two different methods for the determination of
the renormalization constants are presented in \cite{philipp_renorm} which only agree after a chiral extrapolation of the results
is performed. At the same time, the ratio $Z_A/Z_V$ determined from
the values in \cite{philipp_renorm} is almost identical for both methods used
and also stable under chiral extrapolation of the results. 

In our determination of the axial charge from run C, we
encounter what we suspect to be large finite volume effects. Notice that the value of $Z_V$ obtained from the
nucleon three-point functions might be plagued by the same effects. As we cannot calculate $Z_A$ from baryon three-point functions, we therefore always use
the ratio $Z_A/Z_V$ from \cite{philipp_renorm}. In the next section, we
discuss in detail which ratios we measure on the lattice to obtain the
renormalized axial charge $G_A$.

\section{Nucleon axial charge from dynamical CI fermions}

The usual approach \cite{Sasaki:2003jh,Yamazaki:2008py} is to extract the nucleon axial charge
from ratios of $G_A$ over $G_V$
\begin{align}
G_A&=\frac{Z_A}{Z_V}\frac{T_A^3(t,t^\prime)}{T_V(t,t^\prime)}\ ,
\label{simple_ratio}
\end{align}
using single correlation functions built from either smeared quarks or gauge
fixed box or wall sources. This approach has the advantage that some of the systematic errors entering
the lattice determination will cancel.

\begin{figure}[bt]
\begin{center}
\includegraphics[width=7cm,clip]{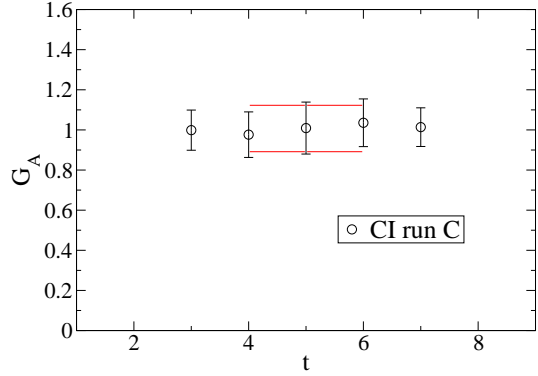} 
\caption[Illustration of a typical plateau observed with our variational
  basis for a source-sink separation of $\approx 1.2\,\mathrm{fm}$]{ Example plot to illustrate typical plateaus observed with our variational
  basis for a source-sink separation of $\approx 1.2\,\mathrm{fm}$. The data
  is from run C.
}
\label{bh_ga}
\end{center}
\end{figure}

We instead use the variational method, which is commonly used to extract
ground and excited state masses. It is based on a correlation matrix $C_{ij}(t)=\langle O_i(t) O^\dagger_j(0) \rangle$
where $O_i(t)$ are operators with the quantum numbers of the state of
interest. The eigenvalues $\lambda_i$ of the generalized eigenvalue problem
$C(t)v_i=\lambda_iC(t_0)v_i$ may be shown to behave as $\lambda_i(t)\propto\mathrm{e}^{-tE_i}\left(1+\mathcal{O}\left(\mathrm{e}^{-t\Delta
  E_i}\right)\right)$, where $E_i$ is the
energy of the $i$-th state. The approach may be generalized to three point
functions. Following \cite{Burch:2008qx}, we obtain an expression for $G_A$:
\begin{align}
G_A&=\frac{Z_A}{Z_V}\frac{\sum_i\sum_j\psi_i^{(k)}T_A^3(t,t^\prime)_{ij}\psi_j^{(k)}}{\sum_l\sum_m\psi_l^{(k)}T_V(t,t^\prime)_{lm}\psi_m^{(k)}}
\ .
\end{align}
Figure \ref{bh_ga} shows a typical plateau for the axial charge of the nucleon from
run C extracted from such a ratio. The horizontal lines denote the
results from a linear fit in the displayed range. Notice that we observe a
plateau in the full range of points we calculated. For all three ensembles, we choose
timeslice $t$ = 9 for the position of the sink. This corresponds to a source-sink
separation of roughly $1.2\,\mathrm{fm}$. For run B, we currently only have
data for insertion timeslices $t^\prime$ from 5 to 9. Instead of assuming that the central
value at 5 is the physical one, we perform a linear fit in the range 5 to 7.

\begin{figure}[bt]
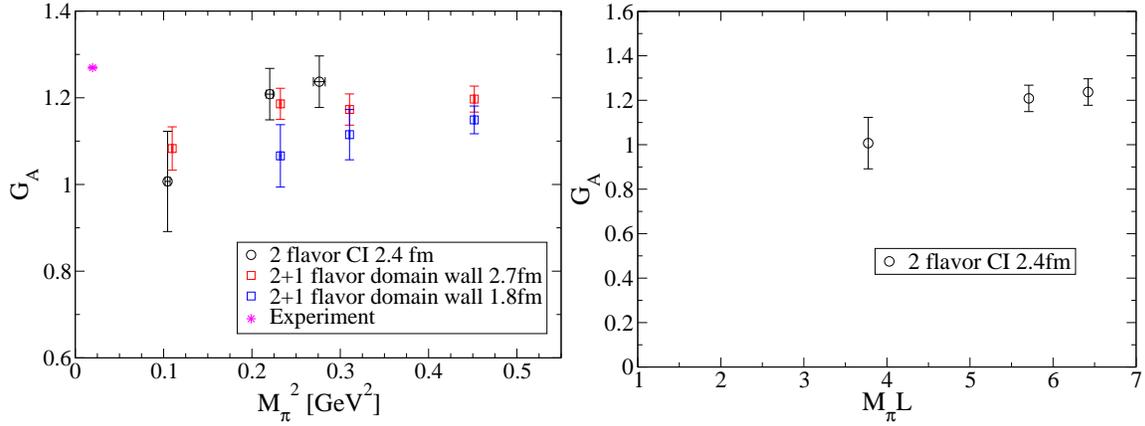

\begin{center}
\includegraphics[height=5.6cm,clip]{ga_over_mpi_compare.eps} 
\includegraphics[height=5.6cm,clip]{ga_over_mpiL_compare.eps} 
\caption[Comparison of our results for $G_A$ to a recent determination using
  domain wall fermions]{ We compare our results for $G_A$ to a recent
  determination from domain wall fermions. Our data is labeled as ``2 flavor
  CI''. Results from 2+1 flavor domain wall fermions are taken from Yamazaki
  et al. \cite{Yamazaki:2008py}. On the left-hand side, we plot the results
  over $M_\pi^2$. Towards lower quark masses finite volume effects are clearly
  visible. On the right-hand side we display our data in units of $M_\pi
  L$.
}
\label{ga_compareplot_nucleon}
\end{center}
\end{figure}

We compare our data to recent results from domain wall fermions
\cite{Yamazaki:2008py} in Fig. \ref{ga_compareplot_nucleon}. The left-hand side plot
shows the results for $G_A$ plotted over the square of the pion mass
$M_\pi^2$. While results at large pion masses lead to values close to the
experimental value, the result from run C deviates substantially from this
behavior. The same is true for the domain wall data and this behavior seems to
be a universal feature associated with finite volume effects
\cite{Yamazaki:2008py,renner_lat09}.  On the right-hand side of the figure we therefore plot the
results for $G_A$ over $M_\pi L$, where $L$ corresponds to the spatial
extent of the lattice. This plot can be directly compared to Fig.
3 of \cite{Yamazaki:2008py}.

Before we move on to calculations for hyperons, let us briefly comment on
the sink-dependence of our results. While results from run A and B are rather
insensitive to the sink location in the region explored (timeslices
9-13), a systematic shift upwards can be observed for run C when reducing the separation
between the source and the sink from 8 to 6 timeslices which corresponds to distances of
$1.2\,\mathrm{fm}$ and $0.9\,\mathrm{fm}$. We want to point out that this does not affect the quality
of the plateau which still stretches over the entire region of insertion
times. Taking a look at the nucleon two point functions,
contributions from excited states to the ground state of the variational
analysis are visible up to timeslice 4. This is an indication that excited
states may indeed be responsible for measuring a larger value of $G_A$ if
excited state contributions are not sufficiently suppressed. With just 50 configurations, the
statistical errors from our preliminary dataset are by far too large to make a
stronger and more quantitative statement.

\section{Hyperon axial charges}

\begin{figure}[bt]
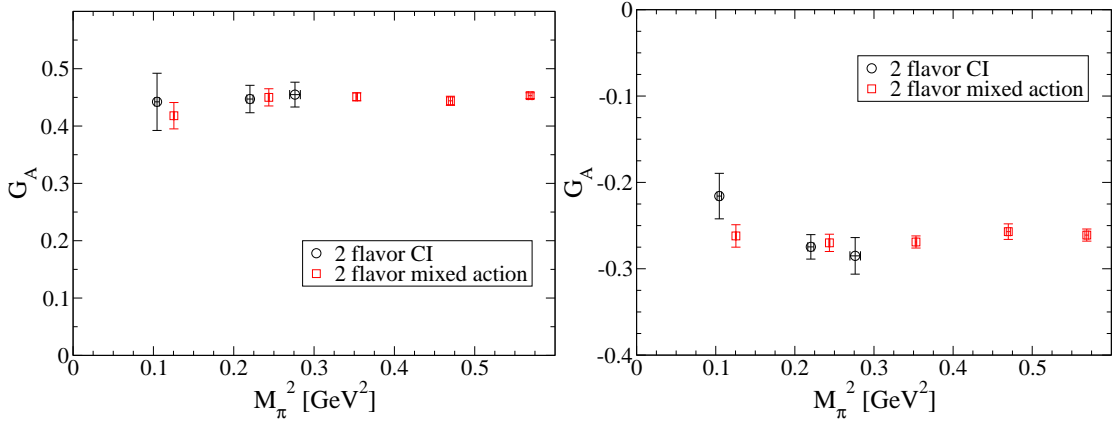

\begin{center}
\includegraphics[width=7.3cm,clip]{sigma_ga_over_mpi_compare.eps} 
\includegraphics[width=7.3cm,clip]{xi_ga_over_mpi_compare.eps} 
\caption{ Results for
  the axial charges of the $\Sigma$ (left-hand side) and $\Xi$ (right-hand side) hyperons
  compared to the mixed action results by Lin and Orginos \cite{Lin:2007ap}.
}
\label{ga_compareplot_sigmaxi}
\end{center}
\end{figure}

In this section we present results for a calculation of hyperon axial
charges. For the $\Sigma$ and $\Xi$ hyperons we adopt the following definitions:
\begin{align}
\left<\Sigma^{+}|A_\mu^3|\Sigma^{+}\right>-\left<\Sigma^{-}|A_\mu^3|\Sigma^{-}\right>=G_{\Sigma\Sigma}\;\bar{u}^\nu\gamma_\mu\gamma_5
u^\nu \ ,\\
\left<\Xi^{0}|A_\mu^3|\Xi^{0}\right>-\left<\Xi^{-}|A_\mu^3|\Xi^{-}\right>=G_{\Xi\Xi}\;\bar{u}^\nu\gamma_\mu\gamma_5
u^\nu \ .
\end{align}
Again, no disconnected contributions appear in the isovector quantities and
the calculation proceeds similar to the nucleon case. In particular, no
additional sequential propagators are needed.

Figure \ref{ga_compareplot_sigmaxi} shows our results for the axial charges
of the $\Sigma$ and $\Xi$ hyperons. We compare our data to \cite{Lin:2007ap}
and we can see a quantitative agreement in the full range of masses. Unlike for
the nucleon, no significant decrease is observed towards the chiral
limit in the plot for the $\Sigma$ (l.h.s.). The plot for the $\Xi$ (r.h.s.)
also shows a nice agreement with the results from \cite{Lin:2007ap}. In this case our value corresponding to the smallest pion
mass shows a slight decrease towards the chiral limit, but the error bars are
large and this may as well be an effect of our limited statistics. Our purely statistical errors on the preliminary dataset of 50
configurations are still large but can be substantially reduced by
using our full statistics. 

\section{\label{ga_outlook}Summary and outlook}

We have presented preliminary results from a calculation of baryon axial charges using a full
variational basis to efficiently suppress contaminations from excited
states. We used a basis of baryon interpolators with different Dirac structures
and two different smearing widths for the quarks. The results are in good
agreement with the literature and we obtain clear plateaus for ratios
calculated with the method of \cite{Burch:2008qx}. Provided the signal for the
states in question is strong enough, this method can also be applied to several
other quantities of interest, among them the axial charge of the delta baryon
and the $N$-$\Delta$ or $\Sigma$-$\Lambda$ transition.

In general, the method we use can also be applied to three-point functions
involving excited states, provided that the signal is good enough to ensure
the necessary separation between the source/sink and the current insertion.

\acknowledgments

The calculations have been performed on the SGI Altix 4700 of the
Leibniz-Rechenzentrum Munich and on local clusters at ZID at the University of
Graz. We thank these institutions for providing support. M.L. and
D.M. are supported by the DK W1203-N08 of the "Fonds zur F\"orderung
wissenschaflicher Forschung in \"Osterreich''. D.M. acknowledges support by
COSY-FFE Projekt 41821486 (COSY-105) and G.E. and  A.S. acknowledge
support by the DFG project SFB/TR-55.

\bibliography{bibtex}
\bibliographystyle{myutcaps}

\end{document}